\documentclass[twocolumn,aps,floatfix]{revtex4}
\usepackage{amssymb} \usepackage{amsmath} \usepackage{amsfonts} \usepackage{amsbsy} \usepackage{fixmath}
\usepackage[T1]{fontenc} \usepackage{pstricks} \usepackage{graphicx}
\usepackage{bbding}
\usepackage{bbm}
\usepackage{bm}
\usepackage[colorlinks=true,citecolor=blue,urlcolor=red]{hyperref}
\usepackage{color}
\usepackage{comment}
\usepackage[normalem]{ulem}
\usepackage{soul}
\usepackage{subfigure}

\begin{document}
\title{Supersolidity of dipolar Bose-Einstein condensates induced by coupling to fermions}

\author{Maciej Lewkowicz,$\,^1$ Tomasz Karpiuk,$\,^2$  Mariusz Gajda,$\,^3$  and Miros{\l}aw Brewczyk$\,^2$ }

\affiliation{
\mbox{$^1$ Doctoral School of Exact and Natural Sciences, University of Bia{\l}ystok, ul. K. Cio{\l}kowskiego 1K, 15-245 Białystok, Poland} 
\mbox{$^2$ Wydzia{\l} Fizyki, Uniwersytet w Bia{\l}ymstoku,  ul. K. Cio{\l}kowskiego 1L, 15-245 Bia{\l}ystok, Poland}
\mbox{$^3$ Institute of Physics, Polish Academy of Sciences, Aleja Lotnik{\'o}w 32/46, PL-02668 Warsaw, Poland} 
}

\date{\today}

\begin{abstract}
We study a mixture of a repulsive 
dipolar condensate and a degenerate Fermi gas in a quasi-one-dimensional geometry. We demonstrate that the presence of fermions, which attract bosons, drastically changes the behavior of the dipolar condensate. For strong enough boson-fermion attraction, a dipolar Bose-Fermi droplet appears in the mixture, and as the attraction becomes stronger, a roton excitation develops in the Bogoliubov spectrum, leading to the formation of a supersolid, and eventually a crystal of isolated droplets. We describe the system  by coupled extended Gross-Pitaevskii (bosons) and Hartree-Fock (fermions) equations.     
We study the excitation spectrum of the system and identify a number of Goldstone and Higgs modes in the supersolid regime. 
%Scaling arguments show that, although the dysprosium atoms are considered to demonstrate the appearance of the supersolid phase, such a phase can be observed with less magnetic atoms.
\end{abstract}

\maketitle

{\it Introduction.--}
The notion of supersolidity was introduced already many years ago \cite{Boninsegni,Penrose56,Gross56,Andreev69,Leggett70}. The supersolid phase displays simultaneously features typical for solids and those for superfluids, which seems to be counterintuitive behavior at the first glance. Although the search for supersolidity was initially focused on solid helium \cite{Chan13}, the first experimental evidence for the existence of a supersolid phase came from the labs working with ultracold atoms. Two ways to supersolidity were realized, the one exploited the long-range, induced by a cavity, interactions between the atoms \cite{Leonard17,Leonard17a}, the other the spin-orbit coupling \cite{Li17}. Soon after, the third method was demonstrated, involving dipolar quantum gases. The coexistence of spatial order and superfluidity, the hallmark of supersolidity, was established in arrays of quantum droplets made of dysprosium or erbium atoms \cite{Tanzi19b,Guo19,Tanzi19a,Natale19,Bottcher19,Chomaz19,Ferlaino21,Bottcher21,Chomaz23}.

In this letter, we propose a route to supersolidity that employs a mixture of bosonic and fermionic ultracold gases. It is essential that the bosonic atoms possess a magnetic dipole moment, although it need not be large. The transition from a Bose-Einstein condensate (BEC) to a supersolid phase is initiated by fermions, which are assumed to attract the bosons. The admixture of fermions weakens the repulsion between bosons \cite{Chin19}. When the boson-fermion attraction is strong enough, the dipolar quasi-one-dimensional bosonic condensate, in a repulsive configuration, becomes unstable. This instability is of the roton kind and triggers the transition of the system to a supersolid phase.

{\it Model.--}
We consider an atomic Bose-Fermi mixture at zero temperature whose many-body wave function is approximated by a product of the Hartree ansatz for bosons and the Slater determinant for fermions. Bosons are described by the condensate wave function, $\psi_B({\bf r},t)$, while fermions are treated individually. A single-particle orbital, $\psi^F_j({\bf r},t)$, the eigenstate of the Schrödinger equation accounting for the bosonic mean-field, is assigned to each fermionic atom (distinguished by the index $j$)
%Since the fermionic sample is spin-polarized, 
We account for the short-range interactions between the bosons themselves and between the bosons and fermions. Fermions are spin-polarized, thus not interact via the s-wave scattering.  The interactions are characterized respectively by the coupling constants $g_B=4\pi \hbar^2 a_B/m_B$ and $g_{BF}=2\pi \hbar^2 a_{BF}/\mu$, related to the scattering lengths $a_B$ and $a_{BF}$, ($\mu=m_B m_F /(m_B + m_F)$ is the reduced mass of the bosonic and fermionic atoms). We further assume that bosons repel each other, while the boson-fermion interaction is attractive. The bosonic and fermionic clouds are confined in axially symmetric, elongated along the $z$ direction, harmonic traps, $V_{B(F)}({\bf r})=\frac{1}{2} m_{B(F)}\left[\omega^2_{B(F)\perp}(x^2+y^2) + \omega^2_{B(F)\parallel} z^2\right]$. Bosonic atoms are assumed to possess magnetic dipoles aligned along the $x$ axis (head-to-head repulsive alignment). The long-range dipolar interactions are therefore given by $V_{DD}({\bf r}) = (\mu^2_{dip}/r^3)\, (1 - 3\, x^2/r^2)$, where $\mu_{dip}$ is the magnetic moment of the bosonic atom.

%Bosonic atoms are assumed to possess a magnetic dipole align along direction perpendicular to the axis of system's symmetry determined by the prolate shape of the  trapping potential (head to head alignment). Bosons are described by the condensate wave function while fermions are treated individually. A  single-particle orbital being the eigenstate of the Schr\"odinger equation accounting for the bosonic mean-field,  is assigned to each fermionic atom. Since the fermionic sample is spin polarized, the  short-range interactions we include are only those acting between bosons themselves and between bosons and fermions. In addition to the dipolar interactions between bosons, we further assume that bosons repel themselves while boson-fermion interaction is attractive. 

Assuming strong transverse confinement, we reduce the system's geometry to a quasi-one-dimensional one. The condensate wave function and fermionic orbitals are in their ground states in the radial directions; therefore, averaging with these density profiles, the one-dimensional dynamics of the system is given by
\begin{eqnarray}
i\hbar\, \partial_t \psi_B &=& \left[-\frac{\hbar^2\, \partial^2_z}{2 m_B} + V_B(z) + g_b\, n_B(z) + g_{bf}\, n_F(z)
\right. \nonumber  \\
&+& \left. \int V_{dd}({z-z^{\prime}})\,n_B(z^{\prime})\,d z^{\prime} \right] \psi_B(z)  \nonumber  \\
i\hbar\, \partial_t \psi^F_j &=& \left[-\frac{\hbar^2\, \partial^2_z}{2 m_F} + V_F(z) + g_{bf}\, n_B(z) \right] \psi^F_j(z)  \,,
\label{1Dequ}
\end{eqnarray}
where $n_B(z)=|\psi_B(z)|^2$ and $n_F(z)=\sum_j|\psi^F_j(z)|^2$ are reduced bosonic and fermionic densities, respectively. The Eqs.~(\ref{1Dequ}),  without dipolar term, were used to study the formation of Bose-Fermi solitons \cite{Karpiuk04}, whose existence has been recently confirmed experimentally \cite{Chin19}. The coupling constants in Eqs.~(\ref{1Dequ}) are rescaled as $g_b = g_B / (2\pi L_{\perp}^2)$ and $g_{bf} = g_{BF} / (2 \pi L_{\perp}^2)$, where $L_{\perp}=\sqrt{\hbar/(m_B \omega_{B\perp})}$ (we assume equal radial length scales for bosons and fermions, i.e., $m_B\, \omega_{B\perp}=m_F\, \omega_{F\perp}$). The dipolar interaction, $V_{dd}(z)$,  splits into $V_{dd}(z)/\mu_d^2=V_{dd}^{sr}(z)+V_{dd}^{lr}(z)$, ($\mu_d = \mu_{dip} / L_{\perp}$), where the attractive short-range part equals $V_{dd}^{sr}(z)=-2/3\, \delta(z)$ and the repulsive long-range part is
\begin{eqnarray}
V_{dd}^{lr}(z) = \frac{-2 \sqrt{a}\, |z| + \sqrt{\pi}\, e^{\frac{z^2}{4 a}}\, (z^2 + 2 a)\,
{\rm{erfc}} (\frac{|z|}{2 \sqrt{a}}) }{8\, a^{3/2}}  \,,  
\label{Vdd}
\end{eqnarray}
where $a=L_{\perp}^2 /2$, and ${\rm erfc}(x)$ function is the complementary error function.
As shown in \cite{Rakshit19b} the beyond mean-filed corrections are not needed to support the self-bound state in one-dimensional case, therefore they are not included in Eq. (\ref{1Dequ}).

{\it Evidence for BEC to supersolid phase transition.--}
We solve the  time-independent version of Eqs. (\ref{1Dequ}) and search for the ground state densities. All the calculations assume the aspect ratio $\omega_{B\perp}/\omega_{B\parallel}=100$.  In what follows, we will use $L_{\perp}$, $\omega_{B\perp}$, $\hbar \omega_{B\perp} L_{\perp}^3$, and $(\hbar \omega_{B\perp}\, L_{\perp})^{1/2}$ as the units of length, frequency, interaction strength $g_B$ and $g_{BF}$, and magnetic dipole moment $\mu_d$, respectively. 
%We demonstrate the transition to the supersolid phase by choosing the strength of repulsion between bosons as $g_B=0.02$ and the magnetic dipole moment as $\mu_d=0.1$. These particular numerical values translate to the value of bosonic scattering length of a few nanometers and the magnetic moment equal to $10\, \mu_B$, provided $\omega_{B\perp}\approx 2\pi \times 14$Hz and $m_B$ equals the mass of $^{162}$Dy atom. 
The number of fermions is  always $10$, but the number of bosons changes.

\begin{figure}[hbt] 
\includegraphics[width=8.cm]{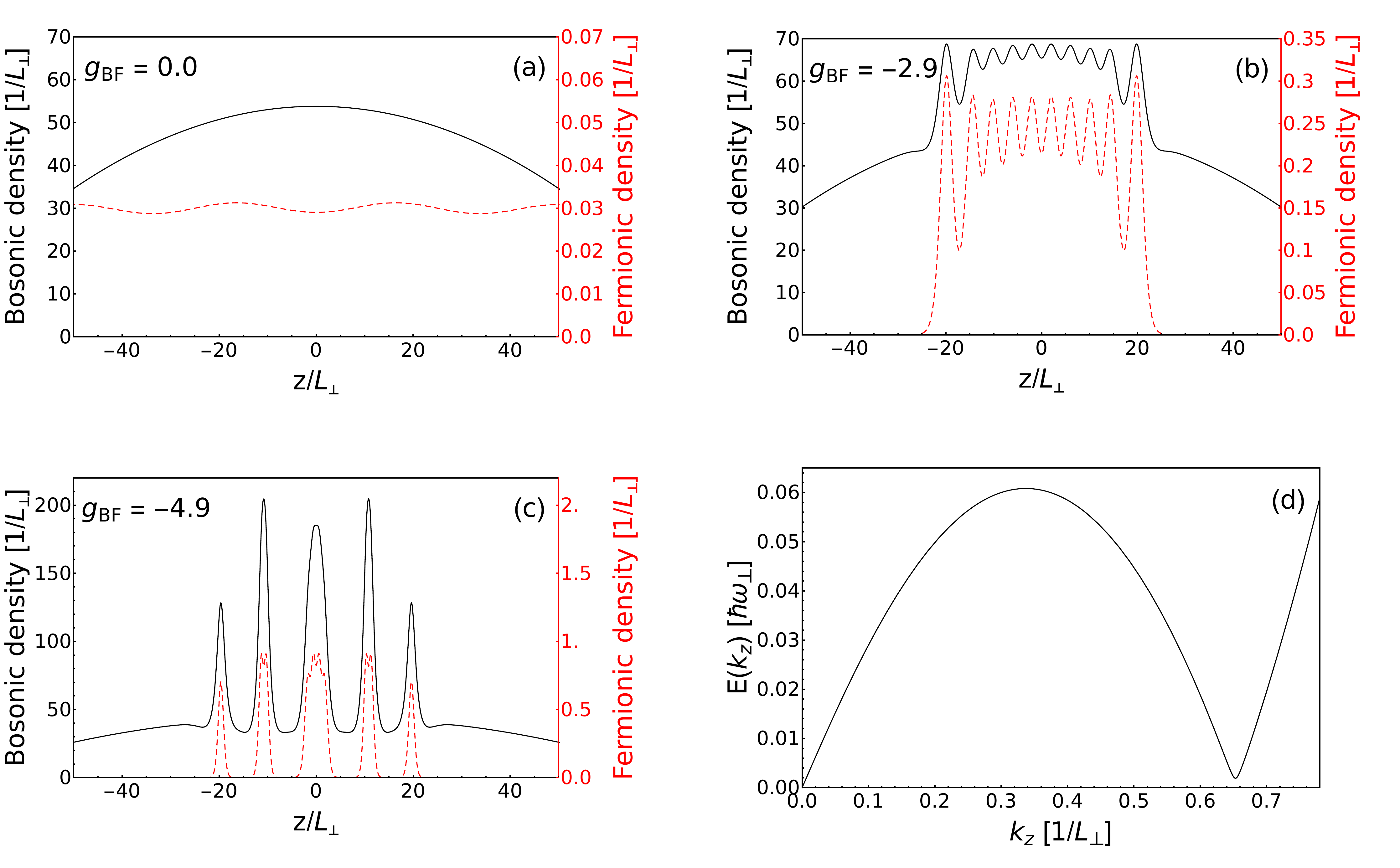}
\caption{(a)-(c) Bosonic (solid line) and fermionic (dashed line) densities for a trapped quasi-one-dimensional system consisting of $N_B=6000$ bosonic and $N_F=10$ fermionic atoms, for different values of boson-fermion attraction: $g_{BF}=0$, $g_{BF}=-2.9$, and $g_{BF}=-4.9$.
(d) Quasiparticle excitation energies for $g_B=0.02$, $\mu_d=0.1$, $-\alpha\, g_{BF}^2=-0.00504251$, ($g_{BF}=-2.9)$,  and the bosonic density $n_B=66$. }
%The roton instability occurs at the critical momentum $k_{cr}=0.65$. This gives the characteristic length equal to about $10$ which is the separation between supersolid density peaks. }
\label{Fig1}
\end{figure}

Fig.~\ref{Fig1} shows the bosonic and fermionic ground state densities for $N_B=6000$ and different values of attraction, $g_{BF}$. Clearly, different stages are visible. For low attraction, the Fermi pressure dominates, causing the bosonic cloud to be hidden inside the fermionic cloud (Fig.~\ref{Fig1}(a)). As the attraction increases, the situation reverses: the fermionic density shrinks, and the fermions become immersed in the bosonic cloud. At a critical value of $g_{BF}$, a dipolar Bose-Fermi (dBF) droplet forms in the mixture. This occurs when the effective interactions between atoms become attractive, analogous to the formation of bright solitons or quantum droplets in Bose-Fermi mixtures, as studied in Ref. \cite{Karpiuk04,Rakshit19a,Rakshit19b}. 
Figure \ref{Fig1}(b) depicts a fully developed dBF droplet at $g_{BF}=-2.9$. It has the shape of a relatively compact and flat fermionic density profile with large-amplitude Friedel-like oscillations. The bosonic cloud follows these oscillations in the region where it overlaps with fermions, while the excess bosons form a broad pedestal.

For even stronger attraction between bosons and fermions, the dBF droplet breaks into five peaks, separated by a distance $d=10$ for both bosons and fermions, as shown in Fig.~\ref{Fig1}(c). There is a transient region in $g_{BF}$ where the Friedel-like oscillations transform into distinct density peaks. Further increasing the attraction sharpens these density maxima, and the bosonic density at different peaks breaks into separate regions.
%The system's density slowly evolves towards the array of well separated  Bose-Fermi solitons (Fig.~\ref{Fig1}(e)).

We interpret the five deep peaks in the density, Fig.~\ref{Fig1}(c), as the appearance of a crystal-like structure resulting from a transition from a single droplet to a supersolid phase. For a fixed number of fermions, $N_F$, and fixed values of $g_B$ and $g_{BF}$, the number of density maxima depends on the number of bosons. It changes from 2 (for $N_B \approx 2000$) to 10 (for $N_B \approx 16000$), which corresponds to the number of fermions  in the system. The maxima in the bosonic density form wells that trap the fermions inside. Since the number of fermions $N_F = 10$ exceeds the number of wells, some of the wells must be occupied by more than one fermion. The fermionic orbitals are delocalized over several bosonic wells.

{\it Roton instability induced by fermions.--} The appearance of a crystal-like structure in a density, as in Fig.~\ref{Fig1}(c), is triggered by a roton instability (RI) developed in the dBF droplet. To show this we further simplify the set of Eqs. (\ref{1Dequ}). This  can be done provided the dynamics of fermions is much faster than that of bosons, i.e. $\hbar/E_F << \hbar/\mu$, where $E_F$ and $\mu$ are the Fermi energy and the chemical potential of bosons, respectively. Under such condition it is possible to eliminate fermionic degrees of freedom, and deal with bosonic system only. However, additional fermion-mediated interactions have to be included. The effective coupling constant for bosons is then changed to \cite{Chin19}, $g_B \to g_B-(3/2) (n_F/E_F)\, g_{BF}^2$,
%\begin{eqnarray}
%g_B - \frac{3}{2} \frac{n_F}{E_F} g_{BF}^2  \,,
%\label{gBnew}
%\end{eqnarray}
where $n_F$ is the density of a uniform fermionic gas. Note that fermion-mediated interactions between bosons are always attractive. Then the uniform system consisting only of bosons is described by the single equation which is of the form of bosonic Eq.~(\ref{1Dequ}) with the interactions replaced by the effective one defined as
$V_{eff}(z)=(g_b - \alpha\, g_{BF}^2 -2/3\, \mu_d^2)\, \delta(z) + \mu_d^2 V_{dd}^{lr}(z)$, where $\alpha = (3/2) (n_F/E_F)/(2\pi L_{\perp}^2)$.
%\sout{defined as} \textcolor{red}{and omitting fermionic term.}

To prove the RI, we analyze the Bogoliubov excitation spectrum, using the approximate formula obtained by eliminating the Fermi component \cite{Giovanazzi04,Santos07}
\begin{eqnarray}
E(k) = \sqrt{\varepsilon(k) \left[ \varepsilon(k) + 2\, n_B \widetilde{V}_{eff}(k) \right] }    \,\,,
\label{excitations}
\end{eqnarray}
where $\varepsilon(k)=\hbar^2 k^2/2m_B$, $n_B$ is the uniform bosonic density, and the Fourier transform of 
%effective interactions 
$V_{eff}(z)$ is \cite{Pawlowski15}
\begin{equation}
\widetilde{V}_{eff}(k) = \left( g_b - \alpha\, g_{BF}^2 - \frac{2}{3}\, \mu_d^2 \right) + \mu_d^2\, \left[f(k^2 a) +1 \right]   \,.
\label{Vkeff} 
\end{equation}
Here, $f(k) = k\, e^{k} Ei(-k)$ where $\rm{Ei}(-k)$ is the exponential integral function. $f(k^2)$ is monotonically decreasing function and inequalities  $0<f(k^2 a)+1 \leq 1$ are always fulfilled. 

The system becomes unstable if elementary excitation energies get imaginary values, i.e. if  the term $g_b - \alpha\, g_{BF}^2 - \frac{2}{3}\, \mu_d^2$ is negative, see Eq. (\ref{Vkeff}),
%To this end it is necessary that the effective interaction overcomes repulsion between bosons and  becomes attractive, $V_{eff}(k)<0$, what might happpen only if the term ($g_b - \alpha\, g_{BF}^2 - \frac{2}{3}\, \mu_d^2$) becomes negative, see Eq. (\ref{Vkeff}). 
%\sout{It is clear that increasing the number of fermions and/or the strength of $g_{BF}$ makes the term $-\alpha\, g_{BF}^2$ more negative}, 
leading to the roton instability in the BEC (RI in binary BEC is discussed in \cite{Bland22,Scheiermann23}). The above analysis is supported by results presented in Fig.~\ref{Fig1}(d)
%Therefore  by increasing the number of fermions and/or the strength of the boson-fermion interaction at fixed $g_b$ and  $\mu_d$, the attraction due to fermions becomes stronger, thus the  magnitude of the negative contribution, $-\alpha\, g_{BF}^2$, is being increased, and a condition for the roton instability in the bosonic component can be reached. This instability is followed by the transition of the system to the supersolid phase. The above analysis is supported by results presented in Fig.~\ref{Fig1}(d)  %The Eq. (\ref{excitations}) can be considered as a two-parameter formula, depending on the bosonic density $n_B$ and the fermionic contribution to bosonic interactions $-\alpha\, g_{BF}^2$. 
for $n_B=66$ and $-\alpha\, g_{BF}^2=-0.00504251$. The critical momentum is $k_{cr} = 0.65$, which corresponds to a characteristic length of approximately 10. Indeed, for a uniform system of size around 50, five pronounced peaks in the density should appear due to the roton instability (see Fig.~\ref{Fig1}(c)). The estimated value of $-\alpha\, g_{BF}^2$ is found to be only 20\% lower than the value obtained numerically.

\begin{figure}[tbh] 
\includegraphics[width=7.0cm]{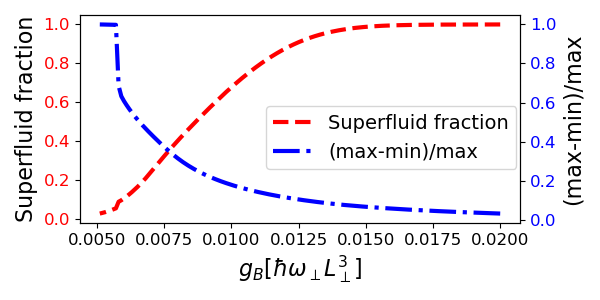} 
\caption{The superfluid fraction (red) and the density contrast (blue)  as a function of boson interaction strength $g_B$. Values of system parameters are: $N_B=3000$, $N_F=10$, $\mu_d=0.06$.}
\label{sf}
\end{figure}

{\it Phases of the system.--} In the following  we study the system if repulsion between bosons is gradually decreased while  boson-fermion attraction is set to  $g_{BF}=-3.8$. The other parameters are selected  such that the supersolid is composed of two maxima only, see caption in Fig.~\ref{sf}.
%i.e. $N_B=3000$, $N_F=10$, $\mu_d=0.06$, and $g_{BF}=-3.8$. 
To  identify  phases of the system we analyse  the superfluid fraction $f_s$, and density contrast, $\delta n$. The upper bound of superfluid fraction,  according to the Leggett criterion \cite{Leggett70}, is given by $f_s < \langle n_B(z) \rangle^{-1} \langle 1/n_B(z) \rangle^{-1}$, where averaging is performed over the size of dBF droplet normalized to one.
%$f_s < \left( \int dz \,n_B(z) \right)^{-1}$ 
A depth  of the density modulations can be measured by the density contrast, $\delta n=(n_{max}-n_{min})/n_{max}$, where $n_{max}$ and $n_{min}$ are values of the density at maximum and  in the middle between the  maxima.

In Fig.~\ref{sf} the superfluid fraction is marked by the red line, while the density contrast by the blue one. Superfluid, present at large values of $g_B$, disappears if   $g_B< 0.005$. Density modulations appear around  $g_B=0.025$, i.e. where the supersolid is formed, and the contrast,  $\delta n$,  grows with decreasing  $g_B$.  Fermions are trapped in the potential formed by the bosonic component, which takes the shape of a broad well with a central bump that grows as $g_B$ decreases, forcing the symmetric fermionic orbitals to 'break' at that point. A minimum, in symmetric orbitals, successively appears at the center: first in the ground state, at $g_B=0.02$, then at  $g_B=0.01$, $g_B=0.0069$, etc. This process leads to a substantial increase in the density contrast.

    In the region of $g_B \lesssim 0.0057$, the self-pinning transition takes place. The effective potential felt by fermions, deep enough to support ten bound states, exhibits two minima separated by high barrier. As the Bose-Bose repulsion decreases from $g_B=0.0057$ to  $g_B=0.005$, the fermions start to occupy either the left or right potential well because parity symmetry is broken by numerical errors when symmetric and antisymmetric states, localized in both wells, are degenerate up to numerical precision. In the self-pinned phase, the contrast jumps to high value  ($\delta n \approx 0.98$) and does not change with $g_B$. A similar pinning transition was observed in a mixture of bosonic and Tonks-Girardeau gases \cite{Busch1} and in two-dimensional droplets with immersed fermionic impurities \cite{Busch2}.

Only at $g_B = 0.005$ does the superfluid fraction practically disappear, and the contrast increases to the value of one, $\delta n = 1$,  which is a signature of the system's crystallization -- formation of  an array of Bose-Fermi droplets with an exponentially small overlap of their bosonic and fermionic wavefunctions.

\begin{figure}[tbh] 
\includegraphics[width=8.0cm]{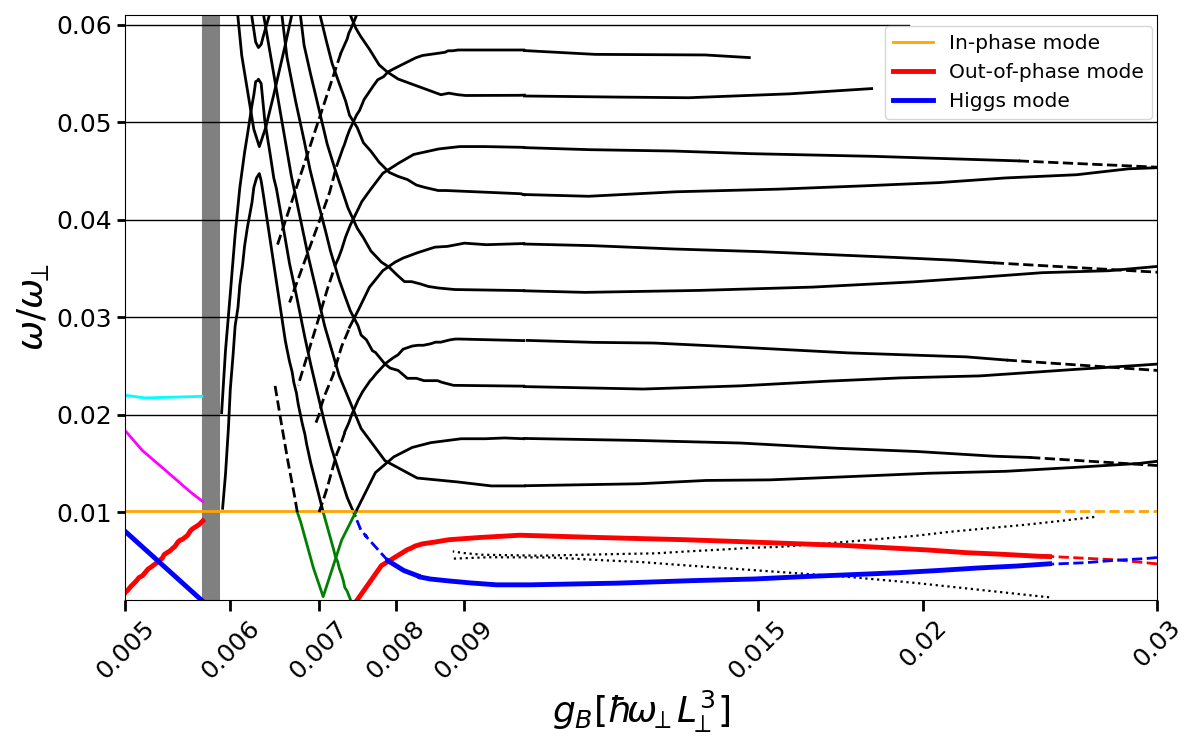}
\caption{Low-energy excitations of bosonic component as a function of the contact repulsion for bosons, in the range of $0.005<g_B<0.03$. Solid red and orange lines show the frequencies of the OPG and IPG modes, respectively. The solid blue line is the Higgs mode. Dotted black lines in the lowest energy sector depict additional very weak modes related to fermions, see SM. } 
\label{spectrum}
\end{figure}

{\it Goldstone modes.--} The low energy excitation spectrum is a very important characteristic of the supersolid.  To excite the system, we weakly perturb the mixture both symmetrically (as in Ref. \cite{Tanzi19b}) and asymmetrically (by a temporal relative energy shift between the two halves of the trapping potential; see the supplemental material (SM) \cite{SM}). We calculate the Fourier transform of the bosonic density, $n_B(z,t)$, in both space and time, $\widetilde{n}_B(k,\omega)$, and look for distinguished frequencies of the function \cite{SM} (compare \cite{Kirkby24}): 
%$\widetilde{n}_B(\omega) = \int \widetilde{n}_B(k,\omega)|\, dk$,
\begin{eqnarray}
\widetilde{n}_B(\omega) = \int |\widetilde{n}_B(k,\omega)|\, dk.  %\,, 
\label{FT1} 
\end{eqnarray}
Positions of local maxima of $\widetilde{n}_B(\omega)$ are excitation frequencies of corresponding modes.
In Fig.~\ref{spectrum} we plot the low-energy excitation spectrum in the relevant range of $g_B$. We present the excitation frequency only if the amplitude of corresponding mode exceeds the background by at least $5\%$.  This is why some lines vanish in the figure, they continue if we lower the $5\%$ criterion (dashed black lines), see SM. 
%\sout{but the price we pay is a false identification of a numerical noise in $\tilde{n}_B(\omega)$  as some eigenfrequencies. A much richer spectrum, where all local maxima of $\tilde{n}_B(\omega)$, including numerical noise, are accounted for, is shown in the SM.} 
To gain more insight into the character of the excitation, we also analyze the time dependence of modes \cite{SM}. Although we focus on the bosonic component, the same results can be obtained studying fermions.

In the harmonic trap, the dynamics of the center of mass (CM) separate from relative excitations. On top of the oscillatory spectrum of equally spaced frequencies
%(horizontal lines in Fig.~\ref{spectrum}), 
the relative excitations are built. Therefore the spectrum repeats itself at every additional CM  excitation quantum. 

For a relatively strong repulsion, $g_B > 0.025$, where the two components form a droplet, the system responds to the external symmetric disturbance mainly through low-energy compressional oscillations with a breathing mode frequency  $\sqrt{5/2}\, \omega_{\parallel}$, similarly to a standard BEC case \cite{Stringari96}. The asymmetric disturbance triggers the  CM oscillations with the axial trap frequency $\omega = \omega_{\parallel}$. 

When the roton instability developes,  around $g_B \approx 0.025$, two  density maxima appear in the bosonic profile -- the supersolid is formed.     
This phase breaks translational symmetry in addition to the U(1) symmetry, which is broken by bosons already  while forming a condensate. Therefore, zero-energy Goldstone modes restoring the broken symmetries should appear \cite{Watanabe11,Watanabe12}. In our case, because of the presence of the external potential, the Goldstone modes are `massive' --  the excitation energies are greater than zero. Such  excitations have already been studied theoretically in ultracold atomic systems \cite{Saccani12,Macri13,Hertkorn19,Zin21}. The system is described by one bosonic and ten fermionic orbitals, which are mutually coupled. Fermionic  orbitals
eventually break translational symmetry (fermions do not have to break this symmetry simultaneously). Therefore, in principle, we might expect multiple Goldstone modes. 
%\sout{The modes corresponding to individual fermionic orbitals should be much weaker than the collective ones.} } 

The CM oscillations of the system can be interpreted as the in-phase Goldstone (IPG) mode, while the relative motion of the superfluid and crystal-like structure is the out-of-phase Goldstone (OPG) mode -- orange and red lines in Fig.~\ref{spectrum}, respectively.  
Dynamics of the OPG mode is characterized by oscillations in the relative heights of the two density maxima.

Simultaneously with the Goldstone modes, the Higgs mode appears as suggested by some relativistic theories. The mode is visible in the Bogoliubov-de Gennes spectrum of dipolar superfluid \cite{Hertkorn19}.  In our case  the Higgs mode manifests itself by the in-phase symmetric oscillations of  amplitudes of density  maxima. In Fig.~\ref{spectrum} the frequencies of the Higgs mode are plotted as solid blue lines. Opposed to the finding of Ref. \cite{Hertkorn19},  the Higgs mode persists in the wide range of the interaction strengths. Evidently interaction with fermions is responsible for this effect.

If boson-boson repulsion  strength, $g_B$, decreases further, the frequencies of OPG and Higgs modes interchange, and at $g_B=0.0075$, the  Goldstone's frequency (red line)  touches zero, while the Higgs's frequency (blue line) reaches the  frequency of the IPG mode. This way the frequencies of the two modes leave the  low frequencies sector of $\omega \in [0,\omega_\parallel]$. Two  other modes (green lines) enter this region instead. 
%\sout{They are visualized in Fig.~\ref{spectrum} by the green lines.}

Rapid  bending of the Goldstone and Higgs modes, observed in the interval  $0.0069< g_B <0.009$, directly relates to the significant sharpening of  bosonic and fermionic density peaks, occurring when the symmetric  fermionic  orbitals start to `break' in the middle, while antisymmetric develop the inflection point, with decreasing $g_B$. The mean-field effective potentials, the system Hamiltonian, and its eigenstates change rapidly then what is visible in  dynamics. Time dependence of the OPG and Higgs modes deviate from their standard behavior as they receive some contribution from the in-phase CM excitations. Similar coupling was observed in \cite{Pylak}. 
%\sout{Green lines in Fig.~\ref{spectrum} indicate this  region.}
The scenario of replacing  modes by another ones at low frequencies, repeats at  $g_B \approx 0.0071$ (green lines).  
%\sout{We are not able to trace these modes at smaller values of $g_B$.}    
A gallery of movies showing the dynamics of the Goldstone and Higgs modes is available at \cite{movies}.

{\it Appearance of low frequency phonon-like modes.--}
The response of the system to a disturbance changes qualitatively for $g_B<0.006$ (vertical black stripe in Fig.~\ref{spectrum}). At such weak boson-boson repulsion, the fermionic cloud  breaks into separate peaks (as in Fig.~\ref{Fig1}(c)), while bosonic cloud is still delocalized. 

The OPG mode appears in the self-pinned phase, its frequency is close to the trap frequency at $g_B \approx0.006$, then it decreases with decreasing $g_B$. Also the Higgs mode revives for $g_B< 0.006$.   There is an energy gap between the Higgs and the Goldstone mode at $g_B\approx 0.006$.

The relative excitations of the crystal-like component -- the crystal phonons, appear at low frequency spectrum in this region. 
Cyan line depicts frequency present in the spectrum of crystal-like peaks inter-distance. The in-phase motion of the two density peaks is realized in the  mode of frequency depicted by magenta line.

Finally, at $g_B\approx 0.005$, the Fourier amplitudes of out-of-phase oscillations of density maxima vanish. This means that superfluid fraction cannot flow between the density maxima -- superfluid simply disappears from the region between them (see Fig.~\ref{sf}). This is the sign of the supersolid to dBF droplets array phase transition. Only the highest frequency phonon mode and the IPG mode persist.

All numerical results presented above were intended to model a mixture of fermionic $^{6}$Li and bosonic $^{162}$Dy atoms. However, since $\mu_{dip} \sim \mu_d\, (m_B^3\, \omega_{B\perp})^{-1/4}$, a scaling of our results is possible. For example, replacing dysprosium by chromium atoms $(m_{B,F} \rightarrow m_{B,F}/3)$ and changing $\mu_d \rightarrow \mu_d/2$ and $\omega_{B\perp} \rightarrow 3^3\,\omega_{B\perp}$ one gets about twice smaller value of $\mu_{dip}$, consistent with the magnetic moment of $^{52}$Cr atoms.

{\it Summary.--}
We have studied the transition to a supersolid phase in a mixture of a dipolar condensate and degenerate fermions within a quasi-one-dimensional geometry. For sufficiently strong boson-fermion attraction, a dipolar Bose-Fermi droplet forms in the system, leading to the development of a roton instability. As a result, a phase with modulated density emerges, indicating the transition to the supersolid regime. The BEC-supersolid transition is further confirmed by analyzing the system's response to axial perturbations, which reveal the appearance of Goldstone modes in the low-energy excitations following the phase transition. For even stronger boson-fermion attraction, the mixture behaves like a solid and, upon disturbance, exhibits phonon-like modes.  Simultaneously, the superfluid motion diminishes and the system finally enters the phase of the array of isolated dBF droplets. Scaling arguments show that the supersolid phase can be observed also with atoms of smaller magnetic moment.

\begin{acknowledgments}
T.K., M.B., and M.G. were supported by the NCN Grant No. 2019/32/Z/ST2/00016 through the project MAQS under QuantERA, which has received funding from the European Union’s Horizon 2020 research and innovation program under Grant Agreement No. 731473. Part of the results were obtained using computers of the Computer Center of University of Bia{\l}ystok.
\end{acknowledgments}

\end{document}

\title{Supplemental materials: Supersolidity of dipolar Bose-Einstein condensates induced by coupling to fermions}

\author{Maciej Lewkowicz,$\,^1$ Tomasz Karpiuk,$\,^2$  Mariusz Gajda,$\,^3$  and Miros{\l}aw Brewczyk$\,^2$ }

\affiliation{
\mbox{$^1$ Doctoral School of Exact and Natural Sciences, University of Bia{\l}ystok, ul. K. Cio{\l}kowskiego 1K, 15-245 Białystok, Poland} 
\mbox{$^2$ Wydzia{\l} Fizyki, Uniwersytet w Bia{\l}ymstoku,  ul. K. Cio{\l}kowskiego 1L, 15-245 Bia{\l}ystok, Poland}
\mbox{$^3$ Institute of Physics, Polish Academy of Sciences, Aleja Lotnik{\'o}w 32/46, PL-02668 Warsaw, Poland} 
}

\maketitle
\begin{figure}[tbh] 
\includegraphics[width=7.0cm]{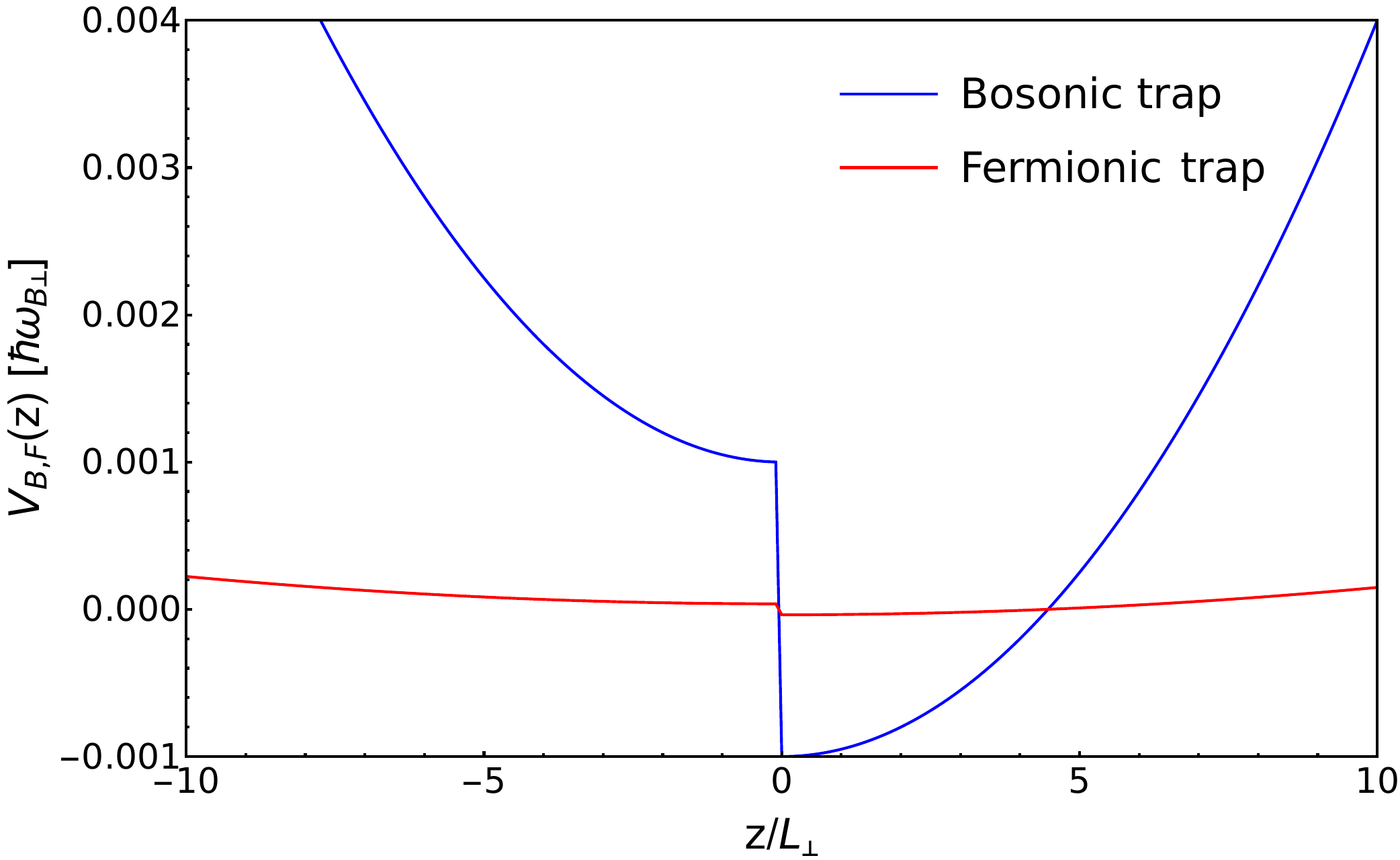} 
\caption{Deformed trap for the excitation spectrum analysis. The blue line corresponds to the bosonic trap $V_B(z)$ and the red line to the fermionic one, $V_F(z)$. }
\label{trap_exc}
\end{figure}

{\it Excitation spectrum.--}
For a given value of the contact interaction parameter $g_B$, we numerically find the ground state of the system in a deformed trap, as shown in Fig.~\ref{trap_exc}. We then instantaneously switch the deformed trap to the unperturbed harmonic trap, evolve the system according to Eq. (1) (main text), and extract the oscillation frequencies using the following method. We calculate the Fourier transform, $\widetilde{n}_B(k,\omega)$, of the bosonic density in both space and time, and then integrate out the momentum dependence
\begin{eqnarray}
&&\widetilde{n}_B(\omega) = \int |\widetilde{n}_B(k,\omega)|\, dk  \,,   \label{FT1} 
%&&\widetilde{n_B}(\omega) = |\int {\mathcal Re}\,[\widetilde{n_B}(k,\omega)]\, dk\,    
%&&\widetilde{n_B}(\omega) = \int |{\mathcal Re}\,[\widetilde{n_B}(k,\omega)]|\, dk  \,, 
\end{eqnarray}
where
\begin{eqnarray}
\widetilde{n}_B(k,\omega) =  \int \int n_B(z,t)\, e^{i \omega t}\, e^{i k z} dz\, dt  \,,  
\label{FT}
\end{eqnarray}
and look for distinguished frequencies of $\widetilde{n}_B(\omega)$.

\begin{figure}[tbh] 
\includegraphics[width=9.0cm]{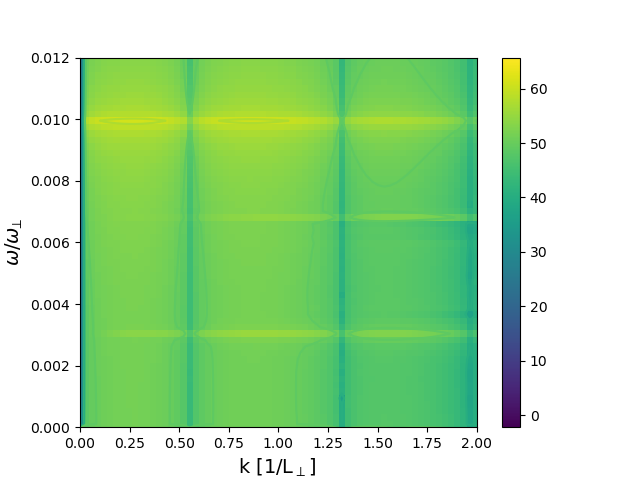} 
\caption{Logarithm of $|\widetilde{n}_B(k,\omega)|\,\, \omega_{\perp}$ as a function of momentum and frequency, for $g_B=0.0085$ and $g_{BF}=-3.8$, showing three distinguished frequencies in the lowest energy sector. They represent (from top to bottom) the in-phase Goldstone mode, the out-of-phase Goldstone mode, and the Higgs mode. }
\label{komega}
\end{figure}

\begin{figure}[tbh] 
\includegraphics[width=5.0cm]{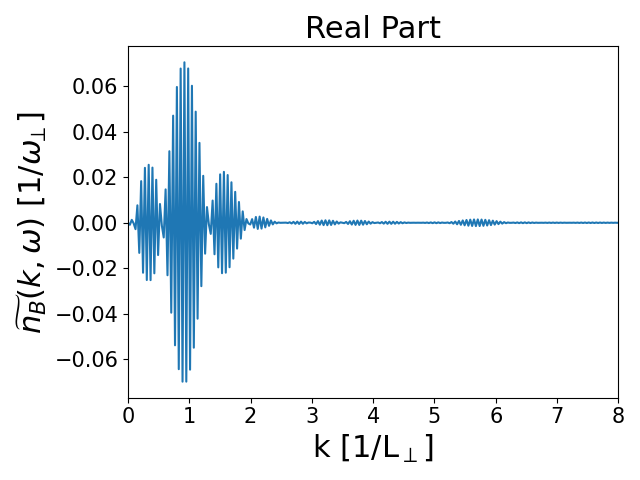} \\
\includegraphics[width=5.0cm]{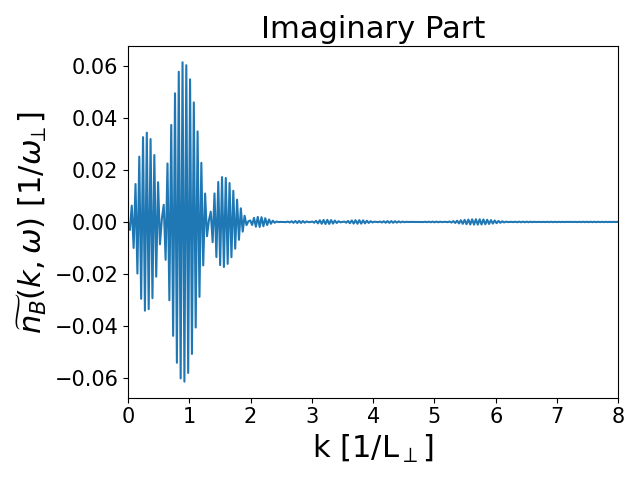} \\
\includegraphics[width=5.0cm]{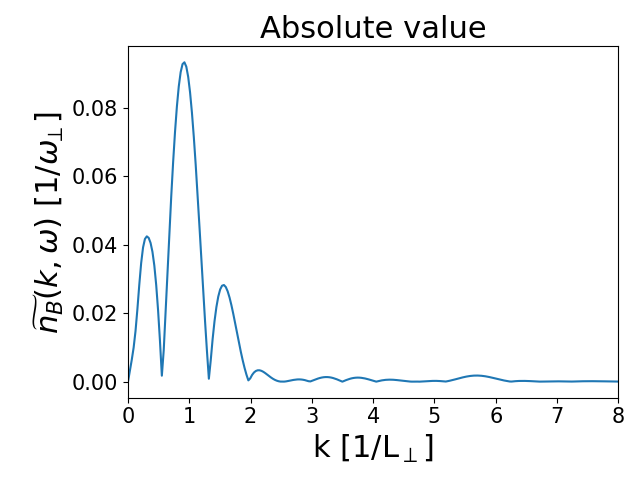}
\caption{${\mathcal Re}\,[\widetilde{n}_B(k,\omega)]$, ${\mathcal Im}\,[\widetilde{n}_B(k,\omega)]$, and $|\widetilde{n}_B(k,\omega)|$ as a function of momentum for the Higgs mode at $\omega=0.003$, for $g_B=0.0085$ and $g_{BF}=-3.8$. }
\label{ReIm}
\end{figure}

\begin{figure}[tbh] 
\includegraphics[width=9.0cm]{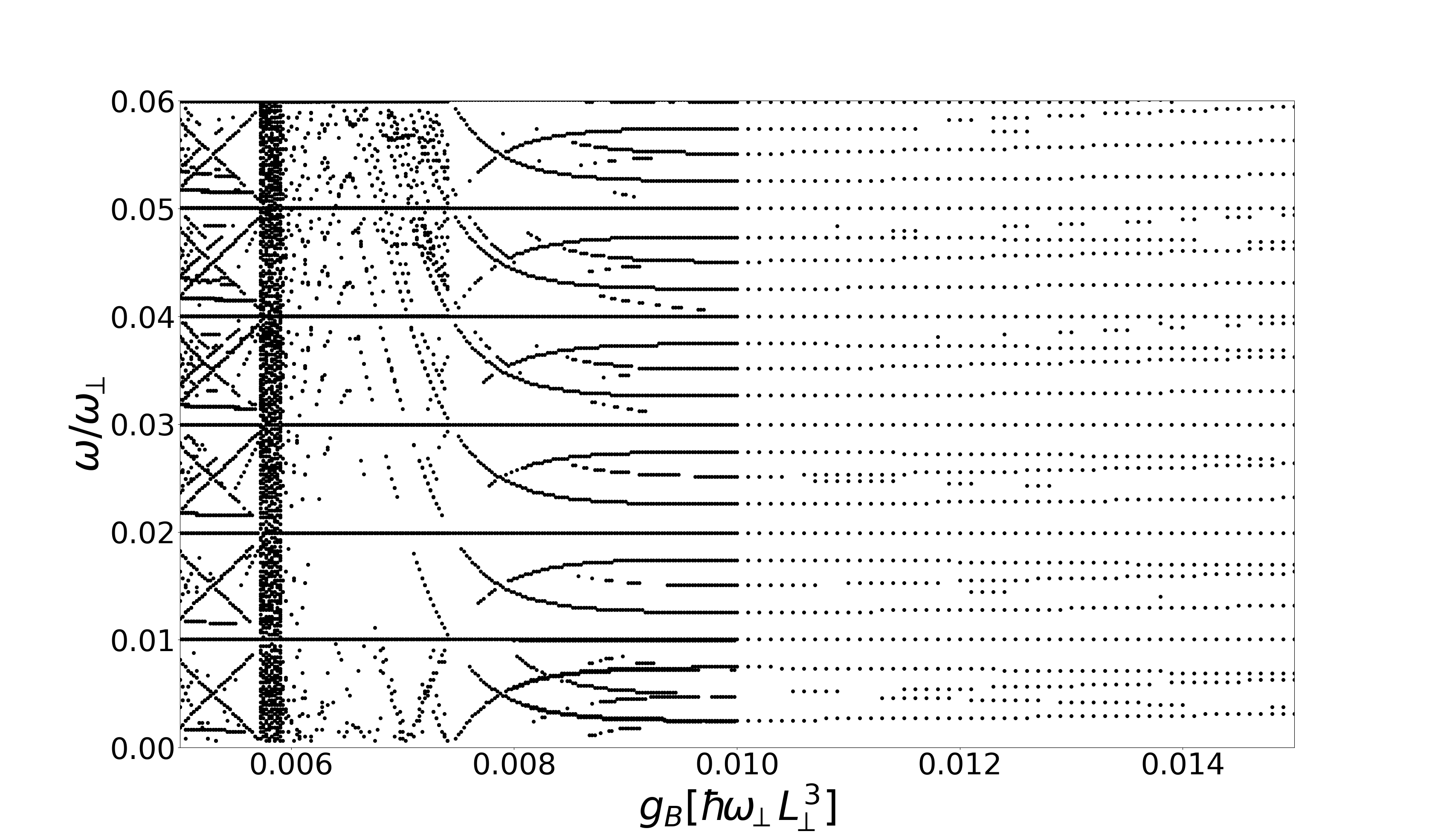} 
\caption{Low-frequency excitation spectrum of bosonic component as a function of the contact repulsion for bosons, in the range of $0.005 < g_B < 0.015$. Data extracted according to Eq. (\ref{FT1}). The visibility of shown frequencies is at least $5$\%. }
\label{excitation_spectrum}
\end{figure}

The modulus of the Fourier transform, $|\widetilde{n}_B(k,\omega)|$, as a function of momentum and frequency is shown in Fig.~\ref{komega} for the contact interaction strength  $g_B=0.0085$ and $g_{BF}=-3.8$. Clearly, three frequencies are distinguished in Fig.~\ref{komega}, visible as horizonatal yellow stripes. They correspond to the Higgs mode (the lowest one), out-of-phase (the middle one), and in-phase (the highest one) Goldstone modes. ${\mathcal Re}\,[\widetilde{n}_B(k,\omega)]$, ${\mathcal Im}\,[\widetilde{n}_B(k,\omega)]$, and $|\widetilde{n}_B(k,\omega)|$ are shown in Fig.~\ref{ReIm} for the Higgs mode at $\omega=0.003$. Nodal points, separating three characteristic groups of momenta, are present (also visible as vertical dark blue lines in Fig.~\ref{komega}). Better insight into the excitation spectrum is gained while looking at $\widetilde{n}_B(\omega)$ (see Eq. (\ref{FT1})) as a function of $g_B$. To find the excitation spectrum we trace the frequency lines which differ from the background by a desired amount.  Fig.~\ref{excitation_spectrum} summarizes the results of such a procedure for 5\% visibility in the range of $0.005 < g_B < 0.0015$. Fig. 3 in the main text was generated based on the above data. We also obtained similar figures for visibility even lower than 5\%. In such cases, however, a false identification of numerical noise in $\widetilde{n}_B(\omega)$ as some eigenfrequencies might occur. We tried to minimize this effect and plotted only the more obvious continuations (denoted by dashed black lines in Fig. 3 in the main text). Even though the visibility contrast was decreased, we were still unable to trace some modes in the range $0.006 < g_B < 0.007$. This is why the left low-energy part of Fig. 3 (main text) is left blank.
We also do not plot weak modes in the range of interactions to the left of the self-pinning phase transition. 

{\it Density oscillations related to the excitation of a given frequency.--}
Dynamic bosonic density related to the excitation of a particular frequency $\omega_j$ is
\begin{eqnarray}
n_B(z,0) + \delta n_B(z,t,\omega_j)   \,,
\label{eq1}
\end{eqnarray}
where $\delta n_B(z,t,\omega_j)$ tells us what is the density disturbance caused by the excitation of frequency $\omega_j$. Since 
\begin{eqnarray}
n_B(z,t) = n_B(z,0) + \delta n_B(z,t)   \,,
\end{eqnarray}
its Fourier transform is given by
\begin{eqnarray}
\widetilde{n}_B(k,\omega) = \widetilde{n}_B(k)\, \delta(\omega) + \widetilde{\delta n}_B(k,\omega)   \,.
\label{eq2}
\end{eqnarray}
Then one has
\begin{eqnarray}
\delta n_B(z,t,\omega_j) &=& \int_{\omega_j-\epsilon}^{\omega_j+\epsilon} \, d\omega  e^{-i \omega t} \int \, \widetilde{\delta n}_B(k,\omega) \, e^{-i k z}    \, dk   \nonumber \\
&=&\int_{\omega_j-\epsilon}^{\omega_j+\epsilon}\, \, d\omega \, e^{-i \omega t} \int \widetilde{n}_B(k,\omega)  \, e^{-i k z}  \, dk \nonumber \\
&-&\int_{\omega_j-\epsilon}^{\omega_j+\epsilon} \, d\omega \, e^{-i \omega t} \, \delta(\omega) \int \widetilde{n}_B(k) \, e^{-i k z} \, dk \nonumber \\
&=& \overline{n_B} (z,t,\omega_j) - n_B(z,0)\, \delta_{0,\omega_j}  \,,
\label{eq3}
\end{eqnarray}
where 
\begin{eqnarray}
\overline{n_B} (z,t,\omega_j) = \int_{\omega_j-\epsilon}^{\omega_j+\epsilon}\, \int \widetilde{n}_B(k,\omega)\, e^{-i \omega t}\, e^{-i k z} d\omega\, dk  \,. \nonumber  \\
\label{eq4}
\end{eqnarray}
If $\widetilde{n}_B(k,\omega) \sim \delta(\omega - \omega_j)$ then
\begin{eqnarray}
\overline{n_B} (z,t,\omega_j) = \left(\int \widetilde{n}_B(k,\omega_j)\, e^{-i k z} dk\right)\,   e^{-i \omega_j t}\,  \,.
\label{eq4a}
\end{eqnarray}
Therefore, for $\omega_j \ne 0$ the second term in Eq. (\ref{eq3}) vanishes and the time dependent density for the excitation mode of frequency $\omega_j$ is determined by the expression
\begin{eqnarray}
n_B(z,0) + \overline{n_B} (z,t,\omega_j)   \,.
\label{eq5}
\end{eqnarray}
Examples of movies showing the time-dependent bosonic densities given by Eq. (\ref{eq5}), representing the Goldstone and Higgs modes in different parameter regimes, are available at \cite{movies}.  Both higher ($g_B = 0.01$) and lower ($g_B = 0.00572, 0.00562, 0.00512$) values of the contact interaction parameter are considered. In addition, we include an animation of the time-dependent density for $g_B = 0.0077$ and $\omega=0.00754$ (dashed blue line in Fig. 3 of the main text), which illustrates the influence of the CM excitation on the Higgs mode.

\begin{figure}[tbh] 
\includegraphics[width=8.5cm]{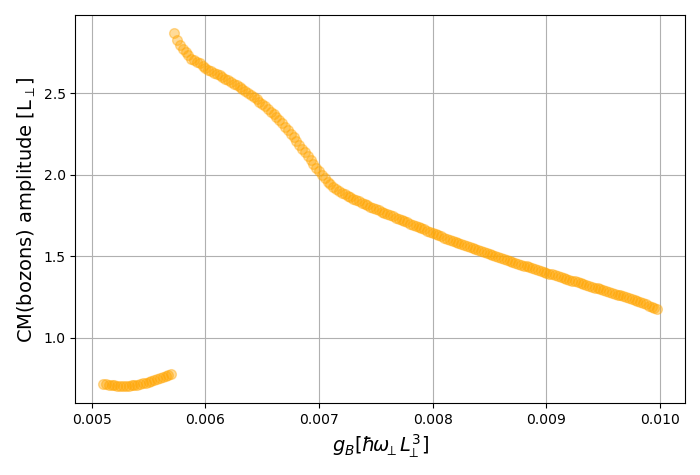} 
\caption{Amplitude of the center of mass oscillations as a function of $g_B$, for $g_{BF}=-3.8$.}
\label{CMamp}
\end{figure}

{\it Center of mass oscillations and fermionic orbitals.--}
As stated in the main text, for $g_B<0.008$, the time dependence of the out-of-phase Goldstone and Higgs modes starts to deviate from their standard behavior, becoming increasingly influenced by the in-phase center of mass (CM) excitations. Fig.~\ref{CMamp} shows the amplitude of the CM oscillations as a function of the contact interaction strength. The amplitude increases as $g_B$ decreases, with regions of both slower and faster growth. This increase in CM amplitude is linked to the shrinking size of the fermionic cloud, which becomes narrower as $g_B$ decreases. This narrowing is a result of the behavior of the symmetric fermionic orbitals, discussed in the main text, which are expelled from the center of the trap as $g_B$ decreases.

\begin{figure}[tbh] 
\includegraphics[width=4.1cm]{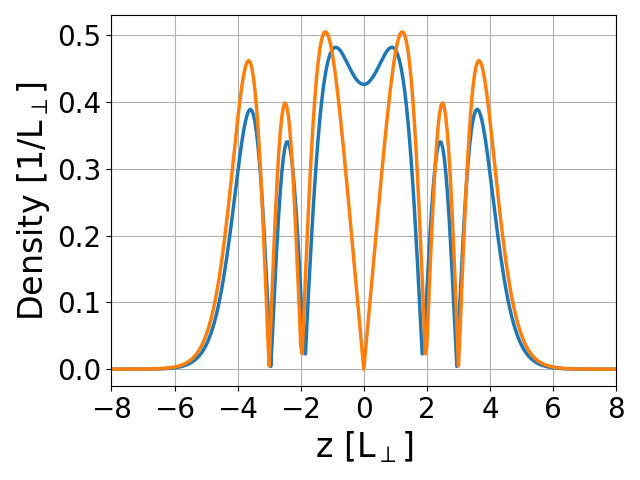} 
\includegraphics[width=4.1cm]{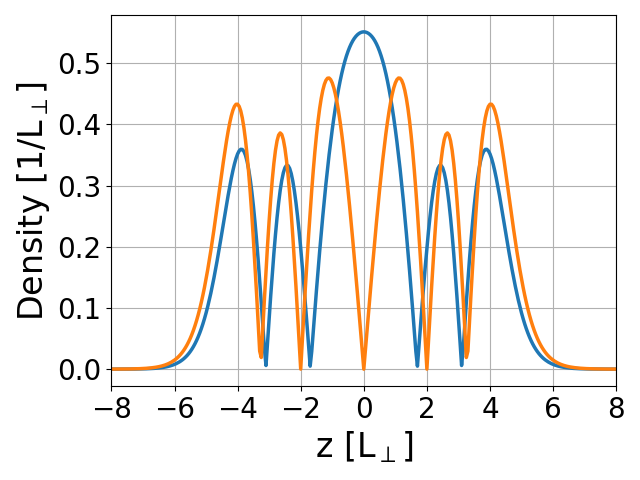} 
\caption{Absolute values of the third (with growing  energy) even-parity (blue line) and the next, odd-parity (orange line) fermionic orbitals, for the contact interaction parameter  above and below the value of interaction strength $g_B \approx 0.007$, where the symmetry of even-parity orbitals is broken. Left frame, $g_B=0.0065$ -- the density corresponding to  the even-parity fermionic orbital  develops a minimum at  the center of the trap.  Right frame, $g_B=0.0077$ -- the density of the same orbital  has maximum at the center. Note the rapid change of the density with varying $g_B$. Odd-parity orbitals are almost intact. }
\label{3even}
\end{figure}

Fig.~\ref{3even} shows the absolute value of the third symmetric fermionic orbital for interactions around $g_B=0.007$, demonstrating the onset of the splitting of this orbital, i.e., the removal of fermionic density from the center of the trap. The bending of the CM oscillation amplitude is clearly observed at  $g_B \approx 0.007$  in Fig.~\ref{CMamp}. Around $g_B=0.006$ the self-pinning phase transition occurs as the fermionic cloud breaks and its size increases.

\begin{figure}[tbh] 
\includegraphics[width=9.0cm]{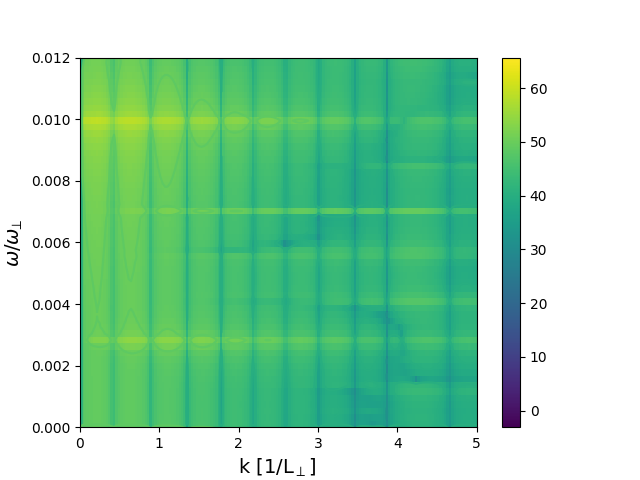}
\caption{Logarithm of $|\widetilde{n}_B(k,\omega)|\,\, \omega_{\perp}$, as a function of momentum and frequency for $g_B=0.013$ and $g_{BF}=-3.8$. Seven distinguished frequencies are clearly visible in the lowest energy sector. Some of them represent the in-phase Goldstone mode ($\omega=0.01$), the out-of-phase (the third frequency line from the top) and the Higgs (the sixth frequency line from the top) modes. The other  modes are related to fermions. }
\label{komega1}
\end{figure}

{\it Small amplitude modes in the lowest energy sector.--}
As shown in Fig. 3 in the main text, there exist also very weak, small amplitude, modes. They are visible in the lowest energy sector, however traces of their `replicas' can be seen also at higher frequencies. Only two of them, the strongest ones, are depicted in figure.  In fact our system is described by eleven orbitals, one bosonic wavefunction and ten fermionic orbitals. 
%The number of fermionc orbitals  equals to the number of fermions in the system. Breaking the translational symmetry of every orbital, should be accompanied by the same number of Goldson modes. Fermionic orbitals, are bound in `bosonic' potential, and their eigenenergies are different, the symmetry can be broken by different orbitals `one by one'.
Since fermions interact with bosons and there are $10$ fermions in the system, one should, in principle (based on the Bogoliubov-de Gennes analysis), expect $11$ times more modes compared to the case of a pure bosonic system. Moreover, the translational symmetry can be broken successively, starting from the lowest energy fermionic orbital, which is most strongly bound to bosons, and ending with the highest energy orbital. 
Amplitudes of oscillations of the modes originated in perturbation of individual fermionic orbitals might be smaller than amplitude of the mode corresponding to the collective excitation of the entire fermionic cloud. 
In Fig.~\ref{komega1} we show evidence for seven modes in the lowest energy sector for $g_B=0.013$ and $g_{BF}=-3.8$, visible as bright horizontal lines. The highest frequency mode is the in-phase Goldstone mode. The Higgs excitation corresponds to the second (from the bottom) frequency and the out-of-phase Goldstone mode is represented by the third (from the top) line in Fig.~\ref{komega1}.

Two frequencies ($\omega=0.00408$ and $\omega=0.00555$) in the middle of the Fig.~\ref{komega1} correspond to some particular points for $g_B=0.013$, on the dotted black lines in Fig. 3 (main text). The dynamics of related modes, obtained from Eq. (\ref{eq4a}) after skipping low momenta components, qualitatively differ. At $\omega=0.00555$, clear out-of-phase Goldstone oscillations are visible, while at $\omega=0.00408$, more complex dynamics is observed.

\begin{figure}[tbh] 
\includegraphics[width=7.5cm]{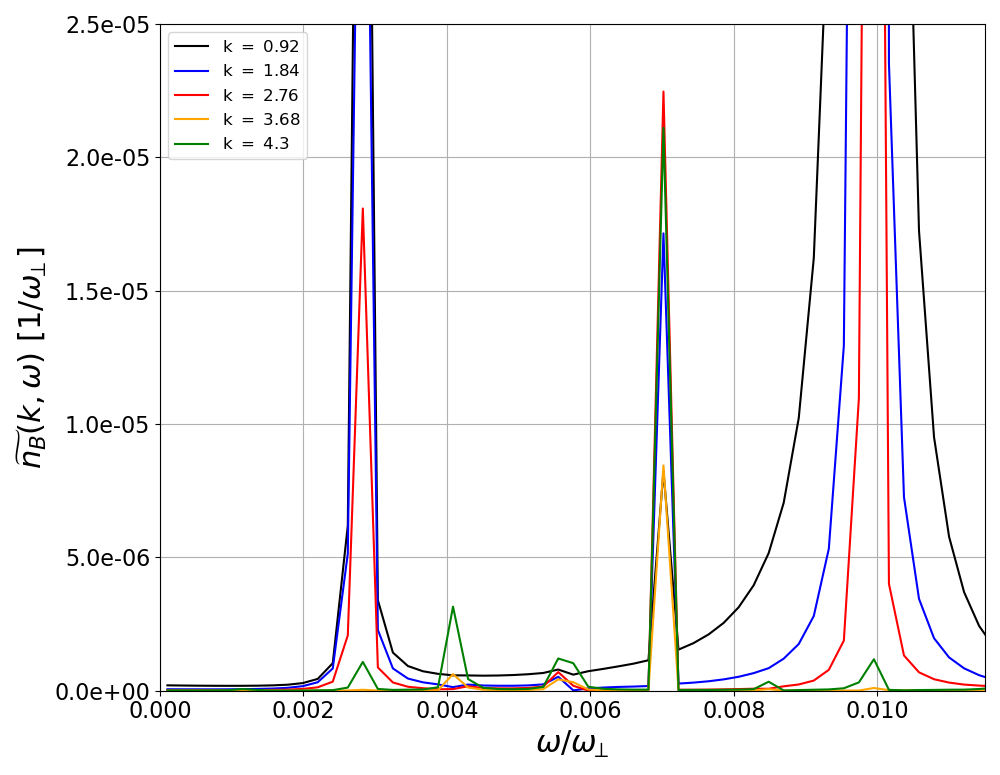}
\caption{Cuts of logarithm of $|\widetilde{n}_B(k,\omega)|\,\, \omega_{\perp}$ (plotted in Fig. \ref{komega1}) for different values of wavevectors, showing that the mode at $\omega=0.00408$ is built of $k>3$,  while  the other one, at $\omega=0.00555$, involves wave vectors corresponding to $k>2$.}
\label{kcuts}
\end{figure}

We want to stress that ignoring small wavevectors $k$ in the Fourier transform $|\widetilde{n_B}(k,\omega)|$ while constructing the two aforementioned eigenmodes is an essential step to observe their regular dynamics. Indeed, the Fourier transform shown in Fig.~\ref{komega1}, as well as cuts through this function at different values of momenta, Fig.~\ref{kcuts}, prove that these modes are built of higher momenta only.

The appearance of additional out-of-phase Goldstone mode in the spectrum is related to the symmetry breaking of the even-parity lowest-energy fermionic orbital in the region of $g_B \approx 0.02$. Breaking symmetry in our trapped system must be understood as the splitting of orbitals into two parts, which manifests as the emergence of a local minimum at the  center. As stated in the main text, the symmetry of the next even-parity fermionic orbital is broken at $g_B = 0.01$. Therefore, one should expect the appearance of another Goldstone mode for $g_B \lesssim 0.01$. Indeed, our analysis for $g_B = 0.009$ confirms the emergence of an additional Goldstone mode. For $g_B = 0.009$, both modes represented by dotted black lines in Fig. 3 (main text) are out-of-phase Goldstone oscillations.

We do not interpret the remaining three modes visible in Fig.~\ref{komega1} since their amplitudes are small, almost at the noise level. Evidently, a complete understanding of the low-energy excitation spectrum requires further detailed analysis, which is beyond the scope of the present study.

